\pdfoutput=1

\documentclass[11pt]{article}

\usepackage[final]{acl}

\usepackage{times}
\usepackage{latexsym}
\usepackage{graphicx}
\usepackage{amssymb}

\usepackage[T1]{fontenc}

\usepackage[utf8]{inputenc}

\usepackage{microtype}

\title{Cross-lingual German Biomedical Information Extraction: from Zero-shot to Human-in-the-Loop}

\author{Siting Liang$^{*}$ \and Mareike Hartmann$^{*,1}$ \and Daniel Sonntag$^{*,2}$ \\
  $^{*}$German Research Center for Artificial Intelligence, Germany \\
  $^{1}$Saarland University, Germany \\
  $^{2}$University of Oldenburg, Germany \\
 \texttt{siting.liang|mareike.hartmann|daniel.sonntag@dfki.de}}
 
\begin{document}
\maketitle
\begin{abstract}
This paper 
presents our project proposal for extracting biomedical information from German clinical narratives 
with limited amounts of annotations. We first describe the applied strategies in transfer learning and active learning for solving our problem. After that, we discuss the design of the user interface for both supplying model inspection and obtaining user annotations in the interactive environment. 
\end{abstract}

\section{Introduction}
Medical information extraction from the large volume of unstructured medical records has the potential to facilitate clinical research and enhance personalized clinical care. Especially the narrative notes, such as radiology reports, discharge summaries and clinical notes provide a more detailed and personalized history and assessments, offering a better context for clinical decision making~\citep{chen2015study,spasic2020clinical}. Name Entity Recognition (NER) task from Natural Language Processing (NLP) studies, have attempted to accurately and automatically extract medical terms 
from clinical narratives~\citep{ sonntag2016clinical, sonntag2019architecture, miotto2018deep, RNNGmodel/lerner2020learning, wei2020study, kim2020ensemble} using annotated clinical text corpora~\citep{mimic/johnson2016mimic,  n2c2/10.1093/jamia/ocz166, miller2019extracting,  biobert/lee2020biobert, clinicalbert/alsentzer2019publicly}. 

The large data collection benefits the research community in developing AI applications in processing medical documents in English~\citep{spasic2020clinical}. However, there are several limitations in improving information extraction from medical records with machine learning methods in other languages, like German in our case: few German annotated datasets are publicly available, and research on non-English medical documents is scarce \citep{StarlingerKittnerBlankensteinLeser+2017+171+179, bronco/kittner2021annotation}. In most cases, domain experts have higher priority commitments and no capacity to annotate large numbers of training examples for use in machine learning applications~\citep{yimam2015interactive}. Our proposed project for extracting medical terms from German clinical narratives with little annotated training data addresses this problem. Two of the most widely studied approaches to this challenge are transfer learning and active learning. 
In transfer learning, models transfer knowledge learned from data-rich languages or tasks to languages or tasks with less or no annotated data~\citep{wang2019cross,lauscher2020zero,xie2018neural, interactive/yuan2019interactive,mbert/pires2019multilingual,mapping/DBLP:journals/corr/abs-1808-09861,plank-2019-neural}. Active learning is an approach to maximize the utility of annotations while minimizing the annotation effort on the unlabeled target data~\citep{chen2015study, miller2019extracting,liu2020ltp, liu2022ltp, chaudhary2019little, shelmanov2019active, zhang2020seqmix, lauscher2020zero}. 
We train a German biomedical NER model building on these two approaches, addressing the following research questions: a) How to transfer knowledge from annotated English clinical narratives corpora to the German NER model? b) In active learning, 
1) What is the minimum amount of annotated samples 
needed for retraining the model? 2) How to evaluate the effectiveness of the query strategies in real-time training (which human (annotator) factors do we have to consider in addition to model performance)?
\begin{figure*}[]
    \centering
    \includegraphics[width=140mm]{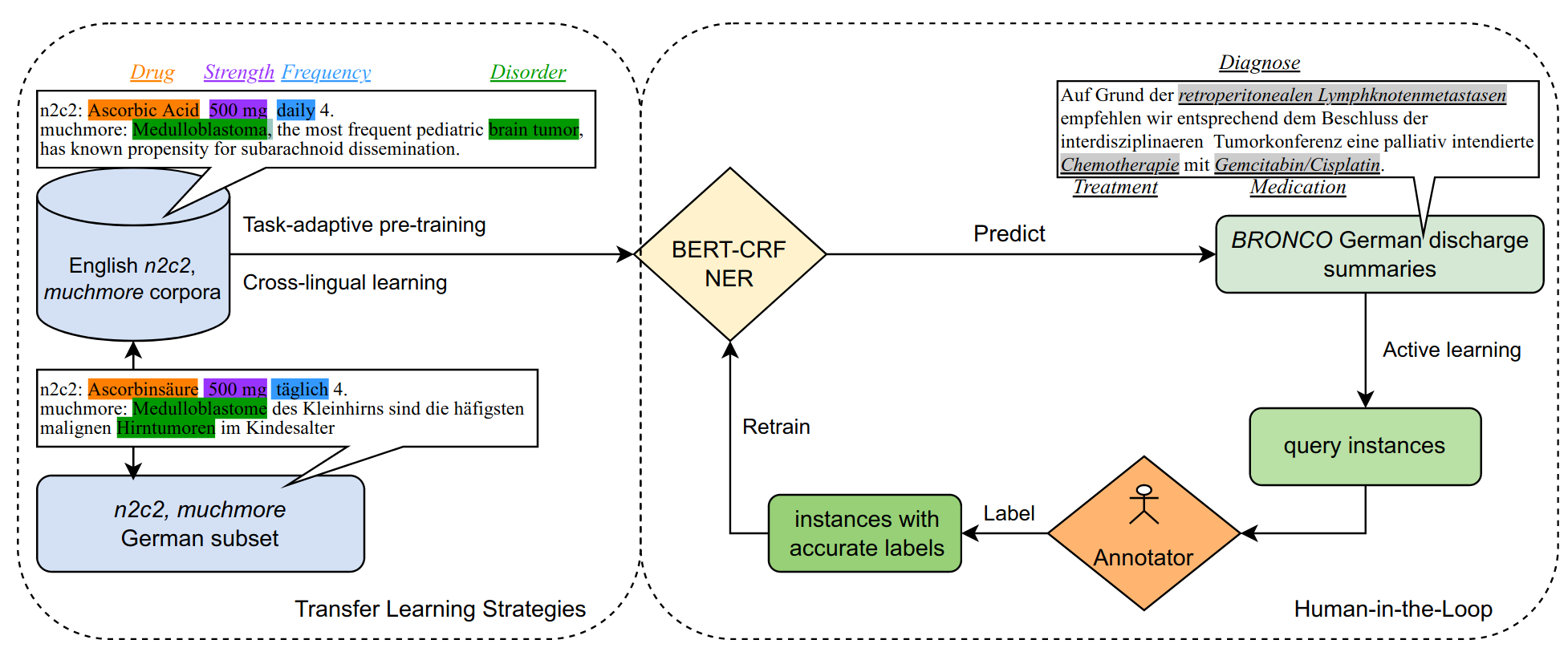}
    \caption{Overview of our 
    research project. We utilize a BERT-CRF architecture as the base NER model. There are three relevant datasets for training and testing which are described in section~\ref{sec:app}. The box on the left side of the figure illustrates the transfer learning strategies: task-adaptive and cross-lingual learning for preparing our base NER model, which are explained in section~\ref{sec:trf}. Human-in-the-loop shown in the right box is the main part of our project, we detail it in section~\ref{sec:hitl}. The design of the user interface for human-in-the-loop is discussed in section~\ref{sec:design}.}
    \label{fig:overview}
\end{figure*}

\section{Approach}
\label{sec:app}
We frame our research problem as NER task for German text in the biomedical domain, and combine transfer and active learning strategies in order to reduce the need for annotated data. Our proposed framework, which is shown in figure~\ref{fig:overview}, is similar to the work of~\citet{chaudhary2019little}. First, a base NER model is pre-trained with English source data using 
transfer learning strategies (see section \ref{sec:trf}). Second, we 
fine-tune the model continuously with annotations in the target language using active learning involving human-in-the-loop (HITL) (see section \ref{sec:hitl}). In contrast to \citet{chaudhary2019little}, we also focus on designing a user interface to obtain human annotations by refining the predictions of the base NER model and pay more attention to the human factors in real-time training (see section \ref{sec:design}).

The NER model used in our framework is a BERT-CRF as in~\citet{liu2020ltp}, which consists of a BERT-encoder~\citep{devlin2018bert} and CRF classifier~\citep{lafferty2001conditional}. The hidden states output from the last layer of the BERT-encoder is fed into the CRF classifier as sequence input to predict the sequence of entity labels. 


\paragraph{Training and Test Data}
Our target task is to extract entities from the German \textit{BRONCO}~\citep{bronco/kittner2021annotation} dataset, which is a collection of discharge summaries with annotated medication terms, i.e. drugs, strength and duration etc, but also other important biomedical information, such as anatomies, diagnosis, and treatments. Two available corpora from the biomedical domain, \textit{n2c2}~\citep{n2c2/10.1093/jamia/ocz166} and \textit{muchmore}~\citep{muchmore/widdows2002using}\footnote{\url{http://muchmore.dfki.de}} have relevant context and useful annotations for our task. Hence, we apply these two datasets for pre-training our base model in order to transfer knowledge to the target task. 
Since neither \textit{n2c2} nor \textit{muchmore} have a matching entity label set for our task, we need to reorganize the training data and define a new joint entity label set. More information about the datasets and defining the entity label set is described in the Appendix~\ref{sec:appendix_a}.
\section{Transfer Learning Strategies}
\label{sec:trf}
We do not train the base NER model from scratch but pre-train it in this part of the work using transfer learning strategies. The aim of pre-training is to transfer the knowledge from annotated English clinical narratives data and German biomedical knowledge to the base NER model for processing the German discharge summaries. 
\paragraph{Task-adaptive Pre-training}
To date, domain-specific language models such as  BioBERT~\citep{biobert/lee2020biobert}, clinicalBERT~\citep{clinicalbert/alsentzer2019publicly}, medBERT~\citep{MEDBERT/rasmy2021med} and BEHRT~\citet{li2020behrt} that are pre-trained on large collections of PubMed abstracts, clinical documents, or electronic health records, are supposed to learn domain knowledge and can directly be applied to 
downstream tasks in the biomedical domain. However, 
domain-adaptive pre-training does not lead to much improvement in the downstream tasks over the general BERT model~\citep{gururangan2020don,Laparra2021ARO}. 
Whereas domain-adaptive pre-training makes use of data from the target domain, task-adaptive pre-training directly uses  unlabeled data from the target task to adapt a pre-trained language model using a language modeling objective, e.g. masked language modeling.

Task-adaptive pre-training with text derived from the specific tasks can further benefit the in-domain LMs performing on these 
tasks in biomedical domain~\citep{Laparra2021ARO}.
In our work, we use the 
available training data for task-adaptive pre-training of the in-domain LMs and find the best pre-trained LM to transfer domain knowledge to our base model. We compare the in-domain LMs and the general BERT model with two criteria: efficiency and accuracy.

\paragraph{Zero- and Few-shot Cross-Lingual Learning}
Cross-lingual learning is a common approach to alleviate the problem of lacking in-language training data but rich annotated English data is available ~\citep{mapping/DBLP:journals/corr/abs-1808-09861, mbert/pires2019multilingual,plank-2019-neural, wang2019cross, zhao2020closer, lauscher2020zero}. The zero-shot setup assumes that no annotated training data is available in the target language, and
multilingual LMs ~\citep{mbert/pires2019multilingual,xlm/lample2019cross} have shown their cross-lingual generalization capabilities in different NLP tasks across ranges of non-English languages~\citep{hu2020xtreme}. However, the impact of linguistic properties of different languages in multilingual models is not yet thorough evaluated~\citep{virtanen2019multilingual}. Research in few-shot transfer learning~\citep{zhang2020seqmix,  chaudhary2019little,lauscher2020zero} has the aim of increasing the performance of the cross-lingual model with only a handful of annotated samples in the target languages. We conduct experiments both in a zero- and few-shot setting to investigate the effectiveness of the multilingual LMs in our task compared to the monolingual in-domain LMs.
Considering that a multilingual model has ten times larger size than the monolingual variants, we also evaluate its efficiency and computation cost both in training and testing time.
\section{Active Learning with Human-in-the-Loop}
\label{sec:hitl}
We apply transfer learning to prepare our base NER model with the source data:~\textit{n2c2} and ~\textit{muchmore} corpora. 
To adapt the NER model to the ~\textit{BRONCO} data, we use active learning to query samples for which we obtain accurate human labels, and improve the accuracy on the target data by retraining the model with the human feedback on these samples. In this part of the work, we do not only analyse the query strategies suitable for the BERT-based deep learning architecture, but also consider the human factors that are expected to strongly affect the human-computer interaction in a real-time training scenario.
\paragraph{Query Strategies from Active Learning} In active learning, we attempt to cope with the problem of little annotation resource by measuring how informative each unlabeled instance is and only labeling the most informative instances with the least effort. The representative query strategies for selecting the samples to label fall into two main categories: uncertainty-based sampling \citep{Lewis1994HeterogeneousUS} and query-by-committee \citep{QBC/10.1145/130385.130417}. When applying active learning to sequence labeling tasks, there are two main issues that we have to address: structured output space and variable-length input. According to results from previous research, sequence level measures are superior to aggregating token-level information for sequence-labeling with CRF models
~\citep{settles2008analysis, chen2015activestudy,shen2017deep, liu2020ltp}. We incorporate the following most representative query methods that are explored in prior work for NER tasks~\citep{settles2008analysis, chen2015activestudy, shen2017deep, chen2017active, siddhant2018deep, shelmanov2019active, chaudhary2019little, griesshaber2020fine, shui2020deep, ren2021survey, liu2020ltp, liu2022ltp, agrawal2021active}, in our experiments:
\begin{itemize}
    \item Lowest Token Probability (LTP) from~\citet{liu2020ltp} as uncertainty-based sampling method;
    \item Batch Bayesian Active Learning Disagreement (BatchBALD)~\citep{houlsby2011bayesian,kirsch2019batchbald} with Monta Carlo Dropout (MC)~\citep{gal2016dropout};
    \item Information Density (ID)~\citep{settles2008analysis, shen2017deep} for addressing the outliers' problem.
\end{itemize}
We detail the mathematical formulations of the query strategies in Appendix~\ref{sec:appendix_c}.

\paragraph{Human Factors}
Collaboration between the human and the model in real-time training is challenging. Most of the previous work in deep active learning only experiment with the query strategies in a simulated scenario~\citep{hil/culotta2005reducing} without measuring the real-time labeling cost and the quality of annotations in practice~\citep{hil/haertel2008assessing,hil/settles2011closing,hil/wallace2018trick, qian2020partner, hil/lertvittayakumjorn2021explanation,hil/wang2021putting, ding2021few, wu2021survey}. Due to the large number of model parameters, deep learning methods can be slow when retraining and force annotator to wait for the next query instance to be labeled~\citep{settles2011theories,arora2009estimating,hil/zhang2019invest}. The more uncertain the predicted labels of the queried instance is, the more corrections are required from the annotator and may lead to inconsistencies among annotators~\citep{chaudhary2019little}. Thus, in addtion to measuring the model performance, we need to consider the following human factors when evaluating the effectiveness of the query strategies in real-time training: i) \textit{annotation workload of each query instance}; ii) \textit{consistency between annotators}.
\section{User Interface Design}
\label{sec:design}
The user interface is critical to the success of the HITL collaboration, as it can affect both user experience as well as the human factors listed above \cite{gajos2008predictability,kangasraasio2015improving}. 
Hence, we aim to implement a user interface informed by recommendations from the human-computer interaction literature, in particular by addressing the four central components of user interfaces for interactive machine learning systems identified by \citet{dudley2018review}. In the following, we describe how we plan to realize each of these components in our system, where possible building on existing interfaces for HITL NER.

\paragraph{Sample review}
The \emph{sample review} component allows the user to assess the state of the model on a given sample. Several available interfaces display the predictions of the current model on the sample to be labeled, with the goal to speed up the feedback assignment step \cite{yang2018yedda,lin2019alpacatag,lee2020lean,trivedi2019interactive}. In contrast, our sample review component focuses on increasing the user's understanding of the state of the model, for example by providing explanations of model predictions along with the predicted labels \cite{stumpf2009interacting, amershi2014power}. To this end, we will experiment with applying gradient- and occlusion-based explainability methods previously studies for sequence classification tasks \cite{atanasova2020diagnostic}.

\paragraph{Feedback assignment}
The \emph{feedback assignment} component allows the user to provide the model with feedback, which can take various forms and constrains the type of interface needed to efficiently collect it. The above mentioned works displaying current model predictions collect label-level feedback by recording the user's binary decision on the correctness of the models suggestions, and a drop-down menu displaying the available label set in case the model prediction is incorrect. \citet{lin2020triggerner} allow the users to mark spans of the input sequence that serve as explanations for a specific prediction. \citet{lee2020lean} additionally collect natural language explanations, using auto-completion to ensure the user provides phrases that can be handled by the system's semantic parser. They find that feedback in the form of important input spans is most efficient, and we plan to focus on this feedback format in combination with label-level annotations.

\paragraph{Model inspection}
The \emph{model inspection} component provides the user with a compact summary of global model performance, e.g. by visualizing performance scores on the validation data. \citet{erdmann-etal-2019-practical} define two complementary evaluation frameworks for active learning models: \emph{Exclusive} evaluation measures model performance on held-out data and indicates how well the model will generalize to additional unlabeled data. \emph{Inclusive} evaluation measures annotation accuracy on the target corpus annotated by user and model jointly. We plan to implement the component such that the user can choose the appropriate metric for the task at hand.


\paragraph{Task overview}
The \emph{task overview} component gives information about additional task-related decisions, e.g. termination criteria, that determine when to stop the annotation process for an optimal cost-benefit trade-off \cite{zhu2007active,laws2008stopping}.

\section{Conclusion}
We have presented our current ongoing work on extracting German biomedical information with a limited number of training sources. To evaluate the applied strategies in an experimental setting, we define our target task based on the training and test data at hand and first engage non-expert annotators in the computer-human interaction. Our long-term goal in this project is to apply our findings and generalize the tool on more diverse types of clinical documents in German. In order to evaluate the effectiveness of HITL in our system from the perspective of other stakeholders, we will be working with physicians in the future.

\section*{Acknowledgements}
The proposed research is funded by the pAItient project
(BMG, 2520DAT0P2).




\bibliography{anthology,custom}
\bibliographystyle{acl_natbib}
\newpage
\appendix
\section{Customized Entity Labels}
\begin{table*}[]
    \centering
    \begin{tabular}{|p{1.8cm}|p{4.5cm}|p{4.5cm}|p{4.5cm}|}
\hline
    Data & \textit{n2c2} & \textit{muchmore}&\textit{BRONCO}\\
    \hline
     Properties& 505 English discharge summaries (303 in training set, 202 in test set), only contain medication annotations, (German \textit{n2c2} contains 6878 annotated German sentences) & Abstracts obtained from PubMed publications of 39 subjects, 7823 in English and 7808 in German, (6374 of them are En-De parallel aligned) &200 deidentified German discharge summaries of cancer patients\\
     \hline
      Relevant Annotations&Drug, Strength, Duration, Route, (ADE) Diseases
Form, Dosages, Frequency&Disorders, Anatomy, Procedures, Chemicals&Diagnosis, Treatments, Medications\\
     \hline
     Utilization&transfer learning &transfer learning&active learning\\
     \hline
     
\end{tabular}
    \caption{Training and test data in our framework. \textit{BRONCO} is the target data in our active learning setting involving human annotators. We pre-train the base NER model with \textit{n2c2} and \textit{muchmore}.}
    \label{tab:data}
\end{table*}
\label{sec:appendix_a}
\textit{muchmore} is a parallel corpus of English-German PubMed abstracts with UMLS (Unified Medical Language System)\footnote{\url{http://umls.nlm.nih.gov}} annotations for training German UMLS vector models. The UMLS term annotations in \textit{muchmore} are about more than thousands of label types but can be concluded into 15 UMLS concept categories. After comparing the entities between \textit{muchmore} and \textit{BRONCO}, four concepts (\textit{Anatomy}, \textit{Procedure}, \textit{Disorder} and \textit{Chemical}) are relevant to our task. We convert two of these labels into entity types required for our target task: (\textit{Procedure} -> \textit{Treatment}), (\textit{Disorder} -> \textit{Diagnosis}). 
Beyond the English \textit{n2c2} corpus, we use the \textit{GERNERMED}~\citep{frei2021gernermed}, which was created by automatically translating a subset of English sentences from \textit{n2c2}. We refer to it as German \textit{n2c2}. The dataset can be used as a resource of parallel English and German sentences. 
Original entity annotations in \textit{n2c2} included: Drug, Strength, Duration, Route, Form, ADE (Adverse Drug Effect), Dosages, Reason and Frequency. German \textit{n2c2} subset does not contain the ADE and Reason labels and focuses on medication administration information, which is more close to our task setting.  We apply the same medication labels as German \textit{n2c2} to our task.
As a result, our NER task contains an entity label set of~$\mathcal{L}=$~\{Drug, Strength, Duration, Route, Form, Dosages, Frequency, Diseases, Anatomy, Treatment, Diagnosis, Chemical\}. Table~\ref{tab:data} shows the details about each dataset and its utilization in our project.

\section{Sequence Labeling with Subtokens}
\label{sec:appendix_b}
The tokenization of the BERT model is based on the WordPiece algorithm~\citep{wu2016google}. Sequence labeling tasks with the BERT model are as a result done at the sub-token level. We follow the instructions of~\citet{devlin2018bert} and only label the first sub-tokens of each word with the BIO tags for training the CRF classifier. The remaining sub-tokens receive the same tags as the [PAD] tokens and are excluded in the loss calculation when predicting over the predefined NER label set. One example is shown in table~\ref{tab:subtoken}
\begin{table}[]
    \centering
    \begin{tabular}{|l|c|c|c|c|c|c|c|c|c|c|c|c|c|}
    \hline
        Token&he & had &not &had& any& di& \#\#ar& \#\#r& \#\#hea& other& than& 1& episode \\ 
        \hline
        Tags &O & O & O & O& O& B-Disorder& x & x & x & O &O& O &O\\
        \hline
    \end{tabular}
    \caption{One example of BIO tagging on subtokens. Only the first subtoken of the entity word "diarrhea" is assigned the entity label and the remaining parts of the word is excluded in the loss computation during training.}
    \label{tab:subtoken}
\end{table}
\section{Mathematical Formulations of Query Strategies}
\label{sec:appendix_c}
For a predicted sequence label $\tilde\mathbf{y}=(\tilde{\mathbf{y}}^1,...,\tilde\mathbf{y}^T)$ and sequence input $\mathbf{x}=(\mathbf{x}_1,...,\mathbf{x}_T)$ with a length of $T$,  we define a query strategy as $\phi\mathbf{x}$ on the predicted sequence of label $\tilde\mathbf{y}$ given the input $x$. 

\begin{enumerate}
    \item \textbf{Lowest Token Probability (LTP)}. The traditional least confidence strategy measuring the uncertainty of $\tilde\mathbf{y}$ by CRF model , i.e. the Viterbi parse~\citep{leastconfidence/culotta2005reducing,settles2008analysis}. Normally, LC is calculated based on the posterior probability $\tilde\mathbf{y}$: 
    \begin{equation}
\phi^{LC}(\mathbf{x}) = 1 - max_{\mathbf{y}^*}P(\mathbf{y}^*|\mathbf{x};\theta)
    \end{equation}
    We adopt the variant of LC proposed by~\citet{liu2020ltp} that measures the least confidence of the tokens in the  in CRF:
    \begin{equation}
\phi^{LTP}(\mathbf{x}) = 1 - min_{\mathbf{y}_i^*\in \mathbf{y}^*}P(\mathbf{y}_i^*|\mathbf{x};\theta)
    \end{equation}
    \item Bayesian Active Learning Disagreement~\citep{houlsby2011bayesian}. \citet{gal2016dropout} proposed \textbf{Monta Carlo Dropout (MC)}for approximating uncertainty in deep learning models. The dropout regularization techniques in deep neural networks are considered as disagreement strategy in bayesian deep learning (BALD)~\citep{houlsby2011bayesian, shen2017deep, siddhant2018deep, kirsch2019batchbald, liu2020ltp,  shui2020deep, ren2021survey}. It estimate the mutual information between the model parameters and model outputs~\citep{ren2021survey}. 
    \begin{equation}
        \phi ^{BALD}(\mathbf{x}) = 1 - \frac{\max(count(\tilde\mathbf{y}^1,...,\tilde\mathbf{y}^T))}{T}
    \end{equation}
    
   $T$ represents the times of performing stochastic forward pass through the network applying different dropout masks that cause individual regularized output $\tilde{{\mathbf{y}}}$ at each time. The average result indicates how confident is the model predicting on the current input instance. \citet{kirsch2019batchbald} adapted BALD in deep learning with batch input: $\phi^{BatchBALD}(x_{i...b})$, where $x_{i...b}$ is a batch of input with size of $b$ .
    \item \textbf{Information Density (ID)}~\citep{settles2008analysis}.  
    To address the problem of sampling outliers, the informativeness of data point $\mathbf{x}$ should be weighted by its similarity to other samples in the original dataset. 
    \begin{equation}
        \phi^{ID}(\mathbf{x}) = \phi^{LTP}(\mathbf{x}) \times (\frac{1}{U}\sum_1^U sim(\mathbf{x},\mathbf{x}^{(\mathbf{u})}))^\beta
    \end{equation}
    The cross-similarity matrix $sim(\mathbf{x},\mathbf{x}^{(\mathbf{u})}))$ among all instances in the dataset can be first computed once and later looked up for each sample in the active learning process. The base query function here $\phi^{LTP}(\mathbf{x})$ is replaceable. $\mathbf{u}$ is the set of samples in the dataset and with the size of $U$. $\beta$ term is a parameter for controlling the importance of the similarity information added to the base informativeness for the given sample~\citep{settles2008analysis}. 

\end{enumerate}

\end{document}